\documentclass[aps,prl,amsmath, amssymb,
nofootinbib,showpacs,twocolumn,tightenlines]{revtex4}
\global\arraycolsep=2pt
\usepackage{bm}
\usepackage{epsfig}

\newcommand{\ben}{\begin{equation*}}
\newcommand{\een}{\end{equation*}}
\newcommand{\bean}{\begin{eqnarray*}}
\newcommand{\eean}{\end{eqnarray*}}

\newcommand{\be}{\begin{equation}}
\newcommand{\ee}{\end{equation}}
\newcommand{\bea}{\begin{eqnarray}}
\newcommand{\eea}{\end{eqnarray}}

\begin{document}
\title{Exact results for Casimir interactions between dielectric
bodies: The weak-coupling or van der Waals Limit}

\author{Kimball A. Milton} 
\email{milton@nhn.ou.edu}
\homepage{http://www.nhn.ou.edu/

\author{Prachi Parashar}
\email{prachi@nhn.ou.edu}

\author{Jef Wagner}
\email{wagner@nhn.ou.edu}
\affiliation{Oklahoma Center for High Energy Physics 
and Homer L. Dodge Department of Physics and Astronomy,
University of Oklahoma, Norman, OK 73019, USA}
\date{\today}

\begin{abstract}
In earlier papers we have applied multiple scattering techniques
to calculate Casimir forces due to scalar fields between different bodies 
described by delta function potentials.  When the coupling to the potentials
became weak, closed-form results were obtained.  We simplify this weak-coupling
technique and apply it to the case of tenuous dielectric bodies, in which
case the method involves the summation of van der Waals (Casimir-Polder)
interactions.  Once again exact results for finite bodies can be obtained.
We present closed formulas describing the interaction between spheres
and between cylinders, and between an infinite plate and a retangular slab
of finite size.  For such a slab, we consider the torque acting
on it, and find non-trivial equilibrium points can occur.
\end{abstract}

\maketitle
The subject of the perturbation of the quantum vacuum energy by material
bodies, the so-called Casimir effect, has a long history
\cite{Casimir:1948dh}.  For only a
limited number of situations, possessing a high degree of symmetry (the
interaction energy between infinite parallel planes \cite{lifshitz}, 
the self-energy of spheres \cite{Boyer:1968uf} and 
cylinders \cite{DeRaad:1981hb}) have exact (numerical) calculations 
been possible.  The experiments
carried out during the last decade or so have typically been for a spherical
surface above a plane surface. (For a review of the experimental
situation, see Ref.~\cite{Onofrio:2006mq}.)
Because the forces for that geometry could
not be calculated exactly, comparision with theory was made using the
proximity force theorem (PFT), which allows one to compute the force between
curved surfaces which are nearly touching \cite{pft}.  
However, there is no well-defined
``proximity force approximation'' that allows one to calculate corrections
to the PFT in powers of the ratio of the separation distance to the radius
of curvature of the surface.  Because the precision of the experiments is
now approaching 1\%, such corrections may become important.

Various interesting developments have improved the theoretical
situation.  For example, there has been notable progress in developing
the numerical Monte-Carlo worldline method of Gies and Klingmuller
\cite{
Gies:2006cq}.   
The difficulty with this
technique lies in the statistical limitations of Monte Carlo methods 
and in the complexity of incorporating electromagnetic boundary conditions.
Optical path approximations have been studied extensively for many years,
with considerable success 
\cite{Scardicchio:2004fy}.  
However, there always remain uncertainties 
because of unknown errors in excluding diffractive effects. 
Direct numerical methods \cite{capasso},  
based on finite-difference
engineering techniques,  may have promise, but the requisite precision
of 3-dimensional calculations may prove challenging.
 A methodology
which is, in principle, exact is the multiple scattering formalism,
which dates back at least into the 1950s
\cite{Wirzba:2007bv,
Bordag:2008gj,
maianeto08,
Emig:2007cf}. 
For more complete references see Ref.~\cite{Milton:2007gy}.

Previous work on this multiple scattering technique, 
which has been brought to a high state of perfection
by Emig et al.\ \cite{Emig:2007cf}, 
has concentrated
on numerical results for the Casimir forces between conducting and
dielectric bodies such as spheres and cylinders.  Recently, we have
noticed that the multiple-scattering method can yield exact, closed-form 
results for bodies that are weakly coupled to the quantum field
\cite{Milton:2007gy}. 
This allows an exact assessment
of the range of applicability of the PFT.  The calculations there,
however, as those in recent extensions of our methodology
\cite{CaveroPelaez:2008tj}, 
have been restricted to scalar fields with $\delta$-function potentials,
so-called semitransparent bodies.  (These are closely related
to plasma shell models.)  Here we remedy that defect by making
the extension of the formalism to electromagnetism, and the bodies
are, correspondingly, 
characterized by a permittivity or dielectric constant $\varepsilon$.
Strong coupling would mean a perfect metal, $\varepsilon\to\infty$, while
weak coupling means that $\varepsilon$ is close to unity.

In this Letter we will briefly review the formalism, and show that
it is precisely equivalent to summing Casimir-Polder or van der Waals
(vdW) potentials.  Exact results have been found in the past in such
summations, for example for the self-energy of a dilute dielectric
sphere \cite{Milton:1997ky} or a dilute dielectric cylinder
\cite{Nesterenko:1997fq}.  Thus it is not surprising that exact results
for the interaction of different dilute bodies can be obtained.
It is only surprising that such results were not found much earlier.
(We note that the additive approximation has been widely used in the
past, for example see Ref.~\cite{Bordag:2001qi}, but here the method is
exact.) We will consider a slab of finite extent above an infinite plane,
and calculate the force and torque on the slab. With the
center of mass fixed, we find generically the shortest side
of the slab aligns with the plate, but for sufficiently square slabs
nontrivial equilibrium points can be found.  We will also compute
the force between spheres and parallel cylinders.  Since the results are
exact, we can quantify the deviation from the PFT.

For electromagnetism, we can start from the formalism of Schwinger
\cite{Schwinger:1977pa}, which is based on the electric
Green's dyadic $\bm{\Gamma}$.
Just as in the scalar case, the vacuum energy for a static
configuration, existing for a time $\tau$, is given by 
\be
E=\frac{i}{2\tau}\mbox{Tr}\ln\frac{\bm{\Gamma}}{\bm{\Gamma}_0},
\ee
and again, in precise analogy with the scalar case, the interaction
in lowest order between two non-overlapping potentials $V_1$ and $V_2$
is
\be
E=\frac{i}{2\tau}\mbox{Tr}\,V_1\bm{\Gamma}_0 V_2\bm{\Gamma}_0.\label{wc}
\ee
Here, the trace is over both vector indices and space-time coordinates,
and $\bm{\Gamma}_0$ is the free Green's dyadic.

Now an easy calculation shows that
\be
E=-\frac{23}{(4\pi)^3}(\varepsilon_1-1)(\varepsilon_2-1)
\int_{v_1}(d\mathbf{r})\int_{v_2}(d\mathbf{r'})\frac1{|\mathbf{r-r'}|^7},
\label{cppot}\ee
where the bodies, which are presumed to be composed of uniform material
filling nonoverlapping volumes $v_1$ and $v_2$, respectively, are 
characterized by
dielectric constants $\varepsilon_1$ and $\varepsilon_2$, both nearly
unity.  This is the famous Casimir-Polder potential \cite{cp}.

Consider first a dilute dielectric slab, a distance $a$ above a dilute
dielectric plate filling the half space $z<0$.  Let the slab have cross
section $A$, and extend from $z=a$ to infinity, as shown in Fig.~\ref{figslab}.
\begin{figure}
\centering
\includegraphics{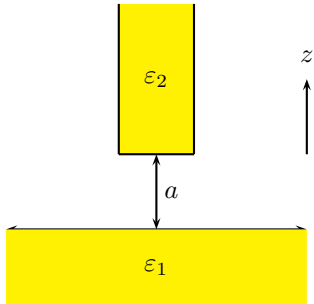}
\caption{\label{figslab} Slab (with dielectric constant $\varepsilon_2$)
of finite cross section $A$ but of infinite extent
in the $z$-direction, a distance $a$ above an infinite plate (with dielectric
constant $\varepsilon_1$) extending to $z=-\infty$.}
\end{figure}
Because the plate is of infinite extent in the $x$-$y$ plane, it is easy to
carry out the integrals in Eq.~(\ref{cppot}), with the result being 
precisely the dilute limit of the Lifshitz formula:
\be
\frac{E}{A}=-\frac{23}{640\pi^2}\frac13
\frac{(\varepsilon_1-1)(\varepsilon_2-1)}{a^3},
\ee
from which the force per area is obtained by taking the negative derivative
with respect to $a$.
There is no correction due to the finite size of the slab.  This is 
presumably a consequence of the fact that for weak coupling, multiple
scattering reduces to the two-scattering approximation.

Instead of integrating over $z$ for the slab, we could have considered
a slab of thickness $dz$ and area $A$, a distance $z$ above
the infinite plate, which has the energy
\be
\frac{dE}A=-\frac{23}{640\pi^2}(\varepsilon_1-1)(\varepsilon_2-1)\frac{dz}
{z^4}.\label{dzslab}
\ee
This allows us to consider an arbitrarily shaped body above the infinite
plate.  For example, we can immediately find the energy of a dilute
sphere of radius $a$, the center of which is a distance $Z$ above
the plate, $Z>a$.  
The energy  is
\be
E=-\frac{23}{640\pi^2}(\varepsilon_1-1)(\varepsilon_2-1)\frac{v}{Z^4}
\frac1{(1-a^2/Z^2)^2},\label{sp-pl-E}
\ee
where $v$ is the volume of the sphere.  When the sphere nearly touches
the plate, $\delta=Z-a\ll a$, we recover the proximity force theorem:
\be
U=-\frac{23}{640\pi^2}(\varepsilon_1-1)(\varepsilon_2-1)\frac\pi3\frac{a}
{\delta^2}.
\label{sp-pl-pfa}
\ee

It is interesting to compare the correction implied by the exact sphere-plate
energy (\ref{sp-pl-E}) to the proximity force theorem result 
(\ref{sp-pl-pfa})
\be
\frac{E}{U}=\left(1-\nu\frac{\delta}{a}\right),\quad d\ll a,
\ee
with $\nu=1.$
This is to be contrasted with the results found by Wirzba \cite{Wirzba:2007bv}
and by Bordag and Nikolaev  \cite {Bordag:2008gj} for
a scalar field with Dirichlet boundary conditions,
$\nu=5/\pi^2-1/3=0.173$; Maia Neto et al.~\cite{maianeto08} 
find for perfectly conducting
boundary conditions for electromagnetic field fluctuations $\nu\approx1.4$.
So our dilute dielectrics have an intermediate behavior.

Now, let us return to the slab geometry, but with finite size in
all directions, and tilted with respect to the infinite plate.
For simplicity, we will consider only a tilt $\theta$ in the $y$-$z$ plane.
See Fig.~\ref{tiltedslab}.
\begin{figure}
\centering
\includegraphics{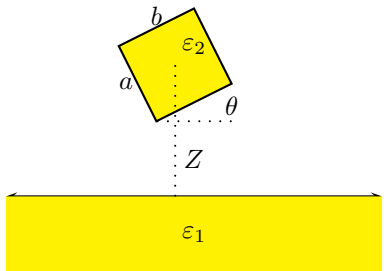}
\caption{\label{tiltedslab} Rectangular solid (with dielectric
constant $\varepsilon_2$) of side $a$, $b$, and $c$ (perpendicular
to the plane, not shown)
 a distance $Z$ above an infinite plate (with dielectric
constant $\varepsilon_1$) extending to $z=-\infty$.
The side $b$ makes an angle $\theta$ with respect to the plate.}
\end{figure}
The center of the body is a distance $Z$ above the plane.
The result of integrating over the body coordinates is
the interaction energy
\be
E=-N\frac{1-\frac{a_+(\theta)}{6Z^2}
-\frac13\left(\frac{a_-(\theta)}{4Z^2}\right)^2}
{\left[1-\frac{a_+(\theta)}{2Z^2}
+\left(\frac{a_-(\theta)}{4Z^2}\right)^2\right]^2},
\label{torque}
\ee
where 
\begin{subequations}
\bea
a_\pm(\theta)&=&a^2\cos^2\theta\pm b^2\sin^2\theta,\\
N&=&\frac{23}{640\pi^2}(\varepsilon_1-1)(\varepsilon_2-1)\frac{v}{Z^4},
\eea
\end{subequations}
and $v$ is the volume of the slab.
This always represents an attractive force between the slab and the
plate.  Most interesting here is the torque exerted about the center
of mass of the slab, $\tau=-\partial E/\partial \theta$. 
Generically, the shorter
side wants to align with the plate, which is obvious geometrically,
since that (for fixed center of mass position) minimizes the energy.
However, if the slab has square cross section, the equilibrium
position occurs when a corner is closest to the plate, also obvious
geometrically.  But if the two sides are close enough in length,
a nontrivial equilibrium position between these extremes can
occur.

\begin{figure}
\includegraphics{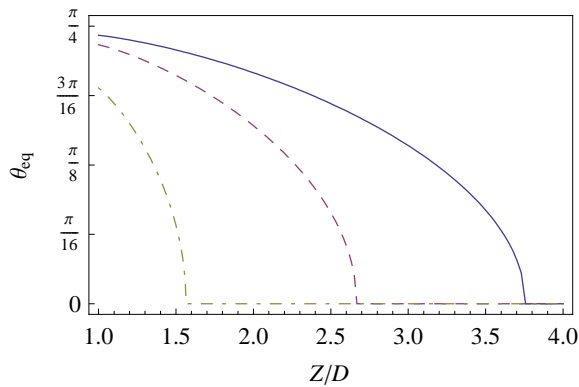}
\caption{\label{figequangle} The stable equilibrium angle $\theta_{\rm eq}$ 
of a slab above an infinite plate for given $b/a$ ratios
0.95, 0.9, and 0.7, respectively given by solid, dashed, and dot-dashed
lines.  For large enough separation, the shorter side wants to face the plate,
but for $Z<Z_0$ the equilibrium angle increases, until finally at 
$Z=D=\sqrt{a^2+b^2}/2$ the slab touches the plate at an angle $\theta=
\arctan b/a$, that is, the center of mass is just above the point of
contact, about which point there is no torque.
}
\end{figure}

The situation is illustrated nicely in Fig.~\ref{figequangle}.  Consider
a slab of aspect ratio $b/a<1$. When one corner of the slab just touches
the plate, the center of mass is $Z=D=\sqrt{a^2+b^2}/2$ and the equilibrium
angle satisfies $\tan\theta=b/a$. 
For large separation, $Z>Z_0$, where
\be
Z_0=\frac{a}2\sqrt{\frac{2a^2+5 b^2
+\sqrt{9 a^4+20 a^2 b^2+20  b^4}}{5 \left(a^2-b^2\right)}},
\ee
the stable equilibrium configuration occurs when $\theta=0$, that is, when the
shortest face $b$ is closest to the plate.  For distances between $Z_0$
and $D$, the stable equilibrium angle is intermediate between 0 and $\tan
b/a<\pi/4$, as shown in Fig.~\ref{figequangle}.
For $a=b$ the stable equilibrium position is always at $\theta
=\pi/4$, and for $b=0$ it is at $\theta=0$.

Obviously the method described here can be used to examine
interactions between bodies of arbitrary shape.  We will content
ourselves here with discussing interactions between cylinders and
spheres.  

First, we remark that it is extremely easy to use this method to
reproduce the results given in Ref.~\cite{Milton:2007gy} 
for parallel cylinders described by $\delta$-function potentials.
It is a bit more  complicated to do the calculation for dielectric
cylinders, of radii $a$ and $b$, separated by a distance $R>a+b$, 
because we have to integrate over the radii of the cylinders.
With the $1/r^7$ van der Waals potentials, the angular integrals can
be carried out explicitly, leaving us with a fairly complicated
function of the radial coordinates on each cylinder.  To proceed,
we expand the integrand of those radial integrals in $r/R$, $r'/R$,
and integrate term by term, $\int_0^a r dr\int_0^b r^{\prime}dr'$.
Once more, it is remarkable that we can explicitly carry out the
sum:
\begin{widetext}
\be 
\frac{E}L=-\frac{23}{60\pi}(\varepsilon_1-1)(\varepsilon_2-1)\frac{a^2b^2}
{R^6}\frac{1-\frac12\left(\frac{a^2+b^2}{R^2}\right)-\frac12
\left(\frac{a^2-b^2}{R^2}\right)^2}{\left[\left(1-\left(\frac{a+b}R\right)^2
\right)\left(1-\left(\frac{a-b}R\right)^2\right)\right]^{5/2}}.
\ee 
\end{widetext}

For two spheres, with center-to-center distance $R$ and radii $a$ and $b$
respectively, the calculation is a bit more complex than for cylinders.
The reason is that even one angular integration leads to elliptic integrals
of complicated argument, so it is difficult to proceed using closed-form
expressions.  Therefore, we expand right at the beginning, before carrying
out any integrations.  We then integrate over polar and azimuthal angles,
and then over the radial coordinates.  We can obtain a power series
expansion in powers of $a/R$ and $b/R$, which it is possible to sum, once
the power series coefficients are identified. 
Again we test the procedure by verifying that we reproduce the results
found previously for semitransparent spheres 
\cite{Milton:2007gy}.
Mathematica is actually able to sum the resulting series, with the following
result for the energy
\begin{widetext}
\be
E=-\frac{23}{1920\pi}\frac{(\varepsilon_1-1)(\varepsilon_2-1)}{R}
\left\{\ln\left(\frac{1-\left(\frac{a-b}R\right)^2}{1-
\left(\frac{a+b}R\right)^2}\right)
+\frac{4ab}{R^2}\frac{\frac{a^6-a^4b^2-a^2b^4+b^6}{R^6}-
\frac{3a^4-14a^2b^2+3b^4}{R^4}+3\frac{a^2+b^2}{R^2}-1}{\left[\left(
1-\left(\frac{a-b}R\right)^2\right)\left(1-\left(\frac{a+b}R\right)^2\right)
\right]^2}\right\}.
\ee
\end{widetext}
This expression, which is rather ugly, may be verified to yield
the proximity force theorem:
\be
E\to U=-\frac{23}{1920\pi}(\varepsilon_1-1)(\varepsilon_2-1)
\frac{a(R-a)}{R\delta^2},\ee $\delta=R-a-b\ll a,b$.
It also, in the limit $b\to\infty$, $R\to\infty$ with $R-b=Z$ held
fixed, reduces to the result (\ref{sp-pl-E}) for the interaction of a sphere
with an infinite plate.

In this Letter we have shown that the general methodology of the multiple
scattering formulation for weak coupling becomes the pairwise
summation of van der Waals energies between the molecules that
make up the dilute dielectrics.  Such summations have previously
been carried out both for the interaction
energy between parallel plates \cite{Schwinger:1977pa}
and the self-interactions of spheres \cite{Milton:1997ky}
and cylinders \cite{Nesterenko:1997fq}, but it was apparently not
recognized that it was easy to obtain exact closed-form energies for
many interesting situations.  Since the calculations here refer to
electromagnetic field fluctuations, causing forces and torques on
dielectric bodies, these are of far more relevance than our earlier
exact results for weak semitransparent bodies.  These results raise
the intriguing possibility that maybe even in strong coupling, 
for example, for conducting bodies, exact results may be obtainable.
In any case, the results presented here form the basis for a laboratory
to study edge effects and other finite-size phenomena which have proved
elusive in the past.

\begin{acknowledgments}
We thank the US National Science Foundation (Grant No.\ PHY-0554926) and the
US Department of Energy (Grant No.\ DE-FG02-04ER41305) for partially funding
this research.  We thank Archana Anandakrishnan and K. V. Shajesh for extensive
collaborative assistance throughout this project.
We are particularly appreciative of Steve Fulling's discussions of the
``Casimir pistol''  that led us to investigate this subject. 
(See Ref.~\cite{fulling08}.)
\end{acknowledgments}


\begin{thebibliography}{99}


\bibitem{Casimir:1948dh}
  H.~B.~G.~Casimir,
 Kon.\ Ned.\ Akad.\ Wetensch.\ Proc.\  {\bf 51}, 793 (1948).

\bibitem{lifshitz} E. M. Lifshitz, Zh. Eksp. Teor. Fiz. {\bf 29}, 94 (1956).

\bibitem{Boyer:1968uf}
  T.~H.~Boyer,
  Phys.\ Rev.\  {\bf 174}, 1764 (1968).


\bibitem{DeRaad:1981hb}
  L.~L.~DeRaad, Jr., and K.~A.~Milton,
  Ann.\ Phys.\ (N.Y.) {\bf 136}, 229 (1981).


\bibitem{Onofrio:2006mq}
  R.~Onofrio,
  New J.\ Phys.\  {\bf 8}, 237 (2006).

\bibitem{pft} J. Blocki, J. Randrup, W. J. \'Swi\c{a}tecki, and C. F.
Tsang, Ann.\ Phys.\ (N.Y.) {\bf 105}, 427 (1977).


\bibitem{Gies:2006cq}
  H.~Gies and K.~Klingm\"uller,
  Phys.\ Rev.\  D {\bf 74}, 045002 (2006).



\bibitem{Scardicchio:2004fy}
  A.~Scardicchio and R.~L.~Jaffe,
  Nucl.\ Phys.\  B {\bf 704}, 552 (2005).
ns, ed.~K. A. Milton, (Rinton Press,

\bibitem{capasso}
A. Rodriguez, M. Ibanescu, D. Iannuzzi, F. Capasso, J. D. Joannopoulos,
and S. G. Johnson, 
Phys.\ Rev.\ Lett.\ {\bf99}, 080401 (2007). 




  






\bibitem{Wirzba:2007bv}
  A.~Wirzba,
  J.\ Phys.\ A  {\bf 41}, 164003 (2008).


 

\bibitem{Bordag:2008gj}
  M.~Bordag and V.~Nikolaev,
  J.\ Phys.\ A  {\bf 41}, 164002 (2008).




\bibitem{maianeto08}
P. A. Maia Neto, A. Lambrecht, S. Reynaud, 
arXiv:0803.2444.
 

\bibitem{Emig:2007cf}
 T.~Emig, N.~Graham, R.~L.~Jaffe and M.~Kardar,
 Phys.\ Rev.\ Lett.\   {\bf99}, 170403 (2007); 
  Phys.\ Rev.\  D {\bf 77}, 025005 (2008);
  T.~Emig and R.~L.~Jaffe,
  J.\ Phys.\ A  {\bf 41}, 164001 (2008).


\bibitem{Milton:2007gy}
  K.~A.~Milton and J.~Wagner,
  Phys.\ Rev.\  D {\bf 77}, 045005 (2008);
  J.\ Phys.\ A  {\bf 41}, 155402 (2008).

\bibitem{CaveroPelaez:2008tj}
 I.~Cavero-Pel\'aez, K.~A.~Milton, P.~Parashar and K.~V.~Shajesh,
  arXiv:0805.2776 [hep-th];
 arXiv:0805.2777 [hep-th];
J. Wagner, K.~A.~Milton, and P.~Parashar, in preparation.


\bibitem{Milton:1997ky}
  K.~A.~Milton and Y.~J.~Ng,
  Phys.\ Rev.\  E {\bf 57}, 5504 (1998).


\bibitem{Nesterenko:1997fq}
K.~A.~Milton,  A.~V.~Nesterenko and V.~V.~Nesterenko,
  Phys.\ Rev.\  D {\bf 59}, 105009 (1999).


\bibitem{Bordag:2001qi}
  M.~Bordag, U.~Mohideen and V.~M.~Mostepanenko,
  Phys.\ Rept.\  {\bf 353}, 1 (2001)
  [arXiv:quant-ph/0106045].

\bibitem{Schwinger:1977pa}
  J.~Schwinger, L.~L.~DeRaad, Jr., and K.~A.~Milton,
  Ann.\ Phys.\ (N.Y.)  {\bf 115}, 1 (1979).


\bibitem{cp}
H. B. G. Casimir and D. Polder, Phys.\ Rev.\ {\bf 73}, 360 (1948).








\bibitem{fulling08} S. A. Fulling, L. Kaplan, K. Kirsten, Z. H. Liu,
and K. A. Milton, arXiv:0806.2468.






\end{thebibliography}
\end{document}